\documentclass[11pt,reqno]{JHEP3}
\usepackage{amsmath,amssymb,longtable,graphicx}

\def\be{\begin{equation}}
\def\beq{\begin{equation}}
\def\ee{\end{equation}}
\def\eeq{\end{equation}}
\def\ba{\begin{eqnarray}}
\def\ea{\end{eqnarray}}

\newcommand{\BZ}{\mathbb{Z}}

\def\one{1\! \vrule height 7.5pt width 0.6pt depth 0pt \vrule height  
0.3pt width 1.1pt depth 0.1pt}
\def\hb{\hat{b}}
\def\hc{\hat{c}}

\newcommand{\mfa}{\mathfrak a}
\newcommand{\mfb}{\mathfrak b}
\newcommand{\mfc}{\mathfrak c}
\def\IP{\relax{\rm I\kern-.18em P}}
\def\CC{{\mathbb C}}

\title{D-brane superpotentials and RG flows on the quintic}  
\author{Marco Baumgartl, Ilka Brunner and 
Matthias R.\ Gaberdiel \\  
Institut f{\"u}r Theoretische Physik, ETH Z\"urich \\
CH--8093 Z{\"u}rich, Switzerland\\ E-mail:
\email{baumgartl@itp.phys.ethz.ch}, 
\email{brunner@itp.phys.ethz.ch}, \email{gaberdiel@itp.phys.ethz.ch}}
\abstract{The behaviour of D2-branes on the quintic under complex
structure deformations is analysed by combining Landau-Ginzburg
techniques with methods from conformal field theory. It is shown that
the boundary  
renormalisation group flow induced by the bulk deformations is 
realised as a gradient flow of the effective space time superpotential
which is calculated explicitly to all orders in the boundary
coupling constant.} 
\preprint{}
\begin{document}

\section{Introduction}

A good understanding of the moduli spaces of string backgrounds is a
central issue in string theory. In many situations of interest the
backgrounds involve D-branes, and then there are two kinds of moduli:
D-brane moduli (that describe the position {\it etc.} of the branes),
and closed string moduli (that parametrise the different closed
string backgrounds). Obviously, these moduli are not independent of
one another, and in particular the D-brane moduli space depends on the
specific closed string background that is being considered.  It is an
important problem (for example in the context of stabilising all
moduli) to understand this dependence in more detail.

In this paper we study this problem for a certain class of
supersymmetric D-branes on the quintic Calabi-Yau threefold. The branes 
we consider wrap rational curves considered in \cite{AK}, and extend along 
the uncompactified dimensions. We will be interested only in the
internal geometry, and hence refer to these branes as D2-branes rather 
than D5-branes. The 2-cycles that are being wrapped can be specified
by linear equations, and at the Gepner point of the quintic they
actually form a complex $1$-dimensional moduli space. This moduli
space can be described explicitly in terms of matrix factorisations of
the associated Landau-Ginzburg model
\cite{Kapustin:2002bi,Brunner:2003dc}. 

As we perturb the closed string theory by a complex structure
deformation, only finitely many D2-branes remain supersymmetric.
In fact, the number of holomorphic spheres has been counted, using
mirror symmetry, in \cite{Candelas:1990rm}. A generic D2-brane will
therefore have to  
adjust itself to the new background. This can be understood from a 
world-sheet point of view following
\cite{Fredenhagen:2006dn}:\footnote{A similar effect has been studied
from an open string field theoretic point of view in
\cite{Baumgartl:2004iy,Baumgartl:2006xb}.} the 
bulk deformation induces a boundary renormalisation group (RG) flow
that drives the brane to one of the allowed discrete
configurations. More specifically, the 
resulting flow is determined by certain bulk-boundary OPE
coefficients. These coefficients are part of the topological sector of
the theory and can thus be calculated  using standard Landau-Ginzburg
techniques following \cite{Kapustin:2003ga}; this allows us to
describe the RG flow completely. As we shall prove fairly
generically, the resulting flow is the gradient flow of the space-time
effective superpotential. 
\smallskip

We can also interpret our calculation as a method to determine
certain contributions of the effective space-time superpotential
exactly. Since the matrix factorisation description applies to 
the full moduli space, we can use it to calculate the open-closed disc
correlation functions with one boundary and one bulk insertion {\em at
every point} in the boundary moduli space. In particular, we can take
the boundary field to correspond to the modulus of the brane moduli
space, and then the correlator can be interpreted as the derivative
(with respect to the boundary modulus) of the full generating function
of the bulk one-point function with an arbitrary number of boundary
fields. By integration we can thus determine from it the exact
contribution to the space-time effective superpotential that is linear
in the bulk modulus. The integral can be performed explicitly, and we
find that the resulting superpotential is a linear combination of
certain hypergeometric functions. Partial results for these
contributions to the effective spacetime superpotential have been
obtained before in \cite{Brunner:1999jq}, and using a complementary
approach more recently in \cite{Ashok:2004xq}. 

Geometrically, the resulting formula has the form of an integral of a 
holomorphic one-form along a path on a Riemann surface
parametrising the possible brane positions. It is therefore
very reminiscent of the analysis of \cite{Aganagic:2000gs} (see also
\cite{Clemens}), who expressed the superpotential for branes wrapping
curves in terms of the Abel Jacobi map by evaluating the holomorphic
Chern Simons action. 
\medskip

Various methods to calculate the effective superpotential have been 
used before in the literature. From a geometric point of view one can
calculate the superpotential as the action of a holomorphic Chern
Simons theory \cite{Witten:1992fb}. For branes wrapping 2-cycles this
simplifies to the integral over certain 3-chains. For compact
Calabi-Yau manifolds (such as the quintic), such integrals are however
difficult to calculate.  

Another approach is to make use of conformal field theory
techniques to evaluate the correlators in perturbation theory.  
This technique was successful in the case of minimal models where all 
correlators could be calculated by making use of the consistency
conditions imposed by the sewing relations \cite{HLL}. For the case of
Calabi-Yau manifolds, however, these consistency conditions by
themselves do not suffice to determine the superpotential
completely. On the other hand, explicit conformal field theory
calculations of correlators are difficult and can at present only be
evaluated at rational points (and to low orders in perturbation
theory). 

For the specific case of the quintic, there are also some explicit
results in the literature. For example, for the Lagrangian (A-type)
branes whose description in terms of conformal field theory was found
in \cite{Brunner:1999jq}, the first few terms of the superpotential
have been determined in  \cite{Brunner:2000wx}, using rational
conformal field theory techniques following \cite{Runkel:1998pm}. The
same problem has also been addressed using Landau-Ginzburg techniques
in \cite{Ashok:2004xq,Hori:2004ja}. More recently, the exact
superpotential has been calculated on the mirror B-side in
\cite{Walcher:2006rs} by guessing the correct Picard-Fuchs equation;
this has also been checked against the instanton expansion of the
A-side that can be interpreted in terms of the counting of holomorphic
discs. Finally, superpotential terms for B-type D-branes described as
pull-backs of sheaves from the embedding $\IP_4$ have been calculated
in \cite{Douglas:2002fr} using linear sigma-model techniques. It was
shown there that the superpotentials obtained in this way capture the
geometry and obstruction theory of the corresponding bundles and
sheaves.  
\bigskip

The paper is organised as follows. In the remainder of this section we
briefly review the geometric description of the quintic and the
relevant family of D2-branes. Section 2 provides background
information about matrix factorisations and their connection to Gepner
models. Moreover the above family of branes is described from a matrix
factorisation point of view. In section 3 we study the RG flow
that is induced by certain complex structure deformations of the
Gepner point. In particular, we find that the RG flow is a gradient
flow and we determine the relevant potential explicitly. In section~4
we explain that the potential can in fact be identified with a certain
contribution to the space-time effective superpotential. Section~5
contains our conclusions. There are two appendices to which some of
technical calculations have been deferred.

\subsection{The model} \label{model}

In this paper we consider D2-branes wrapping holomorphic 2-cycles of
the quintic. Our starting point is the Fermat quintic given by the
following hypersurface in $\IP_4$
\be
\label{Fermat}
x_1^5+ x_2^5 + x_3^5 + x_4^5 + x_5^5=0 \subset \IP_4 \ .
\ee
We are interested in a special family of branes wrapping rational
curves, which has been studied from a mathematical point of view in
\cite{AK} and from a physics point of view in \cite{Ashok:2004xq}, see
\cite{Brunner:1999jq} for earlier work. More concretely, the family of
curves we have in mind is given by 
\be
\label{Katzslines}
(x_1,x_2,x_3,x_4,x_5) =  (u,\eta u,av,bv,cv) \ ,
\quad  \hbox{where} \quad
a^5+b^5+c^5=0 \ .
\ee
Here $a,b,c \in \CC$, $\eta$ is a $5^{th}$ root of $-1$, and $(u,v)$
parametrise a $\IP_1$. The three complex parameters $a,b,c$ are
subject to projective equivalence and the complex equation
in (\ref{Katzslines}), so that the above equations describe a one
parameter family of $\IP_1$'s. In fact there are $50$ such families
since there are $10$ possibilities to pick a pair of coordinates 
that are proportional to $u$, and $5$ choices for $\eta$. These
families intersect along the lines
\be
x_i - \eta x_j =0\ , \quad x_k - \eta' x_l =0\ , \quad x_m =0 \ ,
\ee
where $i,j,k,l,m$ are all disjoint and $\eta$ and $\eta'$ are 
$5^{th}$ roots of $-1$. For example, the set 
\be
x_1-\eta x_2 =0\ , \quad x_3-\eta' x_4=0\ , \quad x_5=0
\ee
describes a particular $\IP_1$ in (\ref{Katzslines}) with $c=0$, 
$a=\eta'$, $b=1$. Likewise, it describes a $\IP_1$ in the
family
\be
(x_1,x_2,x_3,x_4,x_5) = (av, bv, u,\eta' u, cv)
\ee
with $a=1$, $b=\eta$ and $c=0$. Starting from such a configuration,
one can thus move along either of the two families 
of which this $\IP_1$ is part. However, once one has started to move
away in one direction, the other becomes obstructed \cite{AK}. For 
concreteness we shall mainly consider in the following the family of
curves associated to (\ref{Katzslines}) although everything we say
can be easily generalised to the other classes of branes.
 
{}From a conformal field theory point of view, the existence of the
above families of $\IP_1$'s implies that the open string spectrum of
every corresponding brane contains an exactly marginal boundary
operator which we shall denote by $\psi_1$. At the above intersection
points there will be a second exactly marginal operator which we
shall call $\psi_2$ \cite{Ashok:2004xq}. The fact that moving away in
one direction obstructs the other should imply that the effective 
superpotential contains a term of the form  
\be\label{bdlr}
{\cal W}(\psi_1, \psi_2)= \psi_1^3 \psi_2^3 \ . 
\ee
This was argued on physical grounds in \cite{Brunner:1999jq} and later  
confirmed in \cite{Ashok:2004xq}. Recently it was shown in
\cite{Aspinwall:2007cs} that (\ref{bdlr}) is already the full
superpotential for the fields $\psi_1$ and $\psi_2$. We shall 
reproduce this result, using somewhat different methods, at the end of
section~2.
\medskip

The above discussion applies to the Gepner point of the quintic, where
the hypersurface is described by equation (\ref{Fermat}). It is well
known that at a generic point in the complex structure moduli space of
the quintic, there are only discretely many ($2875$) rational
2-cycles; in particular there are therefore no continuous families of
$\IP_1$'s if we perturb the theory away from the Gepner
point. Geometrically, this means that at a generic point in the above
moduli space of branes, the complex structure deformations are
obstructed, as has already been discussed 
in \cite{AK}. From a world-sheet point of view this should therefore
mean that the effective superpotential contains a term of the form  
\be\label{bdlr1}
{\cal W}(\psi_1,\psi_2,\Phi_i)= \psi_1^3 \psi_2^3 + 
\sum_i \Phi_i \, F_i(\psi_1,\psi_2) + \cdots \ , 
\ee
where the $\Phi_i$ describe the different complex structure deformations.  

In this paper we shall mainly consider the special deformations of  
the quintic described by 
\be\label{special}
x_1^5 + x_2^5 + x_3^5 + x_4^5 + x_5^5 + x_1^3 s^{(2)}(x_3,x_4,x_5) 
=0 \ , 
\ee
where $s^{(2)}$ is a polynomial of degree $2$ in $x_3,x_4$ and $x_5$,  
The only curves that survive this deformation are those for which
\be\label{gconstraint}
a^5 + b^5 + c^5 = 0 \qquad \hbox{and} \qquad
s^{(2)}(a,b,c)=0 \ .
\ee
These equations determine a discrete set of points; in fact, counting
multiplicities there are precisely $10$ solutions, as follows from 
Bezout's theorem. 

The deformations (\ref{special}) are special in that the term linear
in $\Phi$ in (\ref{bdlr1}) is independent of $\psi_2$. In this case we
can then determine the function $F(\psi_1,\psi_2)$ exactly, and thus
give a complete description for how the system behaves under the
corresponding bulk perturbation; this will be described in detail in
section~3. As we shall see, the bulk perturbation induces a boundary
RG flow that is the gradient flow of the function $F$; in
particular the solutions to (\ref{gconstraint}) 
are precisely the critical points of $F$.


\section{Branes on the quintic}
\setcounter{equation}{0}

Let us begin by constructing the family of D2-branes
(\ref{Katzslines}) at the Fermat point in the Landau-Ginzburg 
model description.

\subsection{The matrix factorisations description}

At the Fermat point the quintic is described by the Gepner model
corresponding to five copies of the $N=2$ minimal model at $k=3$ 
\cite{Gepner:1987qi,Gepner:1987vz,Greene:1988ut,Witten:1993yc}. 
The branes of interest are B-type branes of this superconformal field
theory. As we shall see, isolated D-branes can be constructed as 
permutation branes in conformal field theory 
\cite{Recknagel:2002qq} (see also \cite{Gaberdiel:2002jr}),
but in order to understand the full moduli space of branes we need
different methods. We shall make use of the fact that in the
associated Landau-Ginzburg (LG) theory, B-type branes 
can be described as matrix factorisations of the LG superpotential $W$  
\cite{Kapustin:2002bi,Brunner:2003dc,Kapustin:2003ga,Kapustin:2003rc,
Lazaroiu:2003zi,Herbst:2004ax} (following unpublished work of
Kontsevich). This formulation is particularly well adapted to the
investigation of brane moduli spaces, as has been observed in various
contexts
\cite{Herbst:2004zm,Ashok:2004xq,Govindarajan:2005im,Brunner:2006tc,%
Jockers:2006sm}.

\noindent At the Gepner point the relevant LG superpotential is    
\be\label{2.1}
W_0 = x_1^5 + x_2^5 + x_3^5 + x_4^5 + x_5^5 \ . 
\ee
A matrix factorisation of $W_0$ is an operator $Q$ that
squares to the LG superpotential
\be
Q^2=W_0 \one \ .
\ee
$Q$ is the boundary part of  the BRST operator, and together with the
bulk BRST charge squares to $0$. In particular, $Q$ is fermionic and 
can be expressed as a linear combination of (non-BRST closed)
fermionic operators $\pi^i$ and their conjugates $\bar\pi^i$,
$i=1,\ldots,n$, that live at the boundary, 
\be\label{Qcliff}
Q= \sum_{i=1}^n \big( \pi^i J_i + \bar\pi^i E_i) \ .
\ee
These fermions form a $2^n$ dimensional representation of the Clifford
algebra 
\be
\{\pi^i,\bar\pi^j\} = \delta^{ij} \ , \qquad 
\{\pi^i,\pi^j\} = \{\bar\pi^i,\bar\pi^j\} = 0  \ . 
\ee
The square of $Q$ is given by
\be
Q^2 = \Bigl( \sum_i E_i J_i\Bigr) \cdot \one
\ee
and hence $Q$ defines a matrix factorisation if
\be\label{factorize}
W= \sum_i E_i J_i \ .
\ee
Turning the argument around, whenever $W$ can be written in the form 
(\ref{factorize}) a suitable matrix factorisation is given by
(\ref{Qcliff}). 
The matrix factorisation description captures all topological aspects
of the corresponding D-branes. For example, one can determine from
$Q$ the topological part of the open string spectrum and the
topological RR charges, {\it etc}. What will be most important for our
purposes is the Kapustin-Li formula \cite{Kapustin:2003ga} that allows
one to calculate bulk-boundary correlators (or boundery three point
functions) exactly. If we denote a topological bulk field by $\Phi$
and the boundary field by $\psi$, then the disc correlator is 
\be\label{KL}
\langle\Phi\, \psi\rangle = 
{\rm Res}\; \Phi 
\frac{ {\rm STr} \left [\,\partial_{x_1}Q \dots \partial_{x_5} Q \,\psi
\right]}{\partial_{x_1}W\dots\partial_{x_5} W}\ ,
\ee
where the residue is taken at the critical points of LG superpotential
$W$. More details about this formula can be found in
\cite{Kapustin:2003ga,Herbst:2004ax}. 

Strictly speaking, to find an LG description of the quintic one has to
consider an orbifold of the theory (\ref{2.1}). This $\BZ_5$ orbifold 
projects onto states with integer $U(1)$ charge in the closed string 
sector. As usual, the consequence for the open string sector is
\cite{Ashok:2004zb,Hori:2004ja,Walcher:2004tx} that we need to specify 
in addition a representation of the orbifold
group on the Chan-Paton labels. The open string spectrum is then given
by the $\BZ_5$ invariant part of the cohomology of the BRST
operator. In the following, the additional representation label will
play no further role, since we will only consider a single D-brane
with an arbitrary but fixed representation label. 
\medskip

The D2-branes of interest correspond to a family of matrix
factorisations that can be constructed as follows, using ideas similar 
to what was done in \cite{Brunner:2006tc,Brunner:2004mt} (see also 
\cite{Hori:2004ja}). We define 
\be\label{ansatz}
J_1 = x_1 - \eta x_2 \ , \qquad J_4 = a x_4 - b x_3 \ , \qquad
J_5 = cx_3 - a x_5  \ , 
\ee
and look for common solutions of $J_1=J_4=J_5=0$ and $W=0$. 
If $\eta$ is a fifth root of $-1$ and $a\neq 0$, we get a solution if  
\be\label{moduli}
a^5 + b^5 + c^5 = 0 \ .
\ee
If this is the case we can use the Nullstellensatz to
write 
\be
W_0 = J_1 \cdot E_1 + J_4 \cdot E_4 + J_5 \cdot E_5 \ ,
\ee
where $E_i$ are polynomials in $x_j$. We then obtain a matrix
factorisation by the procedure outlined above. More specifically, we
introduce $8\times 8$ matrices $\pi^i$ and $\bar{\pi}^i$, $i=1,4,5$,
that form a representation of the Clifford algebra, and obtain a
family  of matrix factorisations $Q(a,b,c)$ 
\be\label{Qref}
Q(a,b,c) = \sum_{j=1,4,5} \left( J_j \pi^j + E_j \bar\pi^j \right) \ . 
\ee
By construction $Q(a,b,c)$ satisfies then 
$Q(a,b,c)^2 = W_0 \cdot \one$. 

Following the geometrical interpretation of matrix factorisations
elaborated in \cite{O,Aspinwall:2006ib,HHP} (see also
\cite{Ezhuthachan:2005jr}) these matrix factorisations provide 
the LG-description of the D2-branes described in section
\ref{model}. Indeed, read as equations in 
$\IP_4$, the equations $J_1=J_3=J_5=0$ describe precisely 
the geometrical lines (\ref{Katzslines}).

The moduli space of such branes has complex dimension one. Indeed, it
is straightforward to see that rescaling $(a,b,c)$ by a common factor 
results in an equivalent factorisation; thus (\ref{moduli}) can be
thought of as an equation in $\mathbb{C}\mathbb{P}^2$, and hence
describes a one-complex-dimensional curve.\footnote{In the above
description we have not treated the three variables $a$, $b$ and $c$
on an equal footing, and hence $a$ could not be zero. It should
be clear, however, that we can also use a different chart in which
$a=0$ is possible. In this way we can obtain a matrix factorisation
associated to $(a,b,c)$ provided that not all three $a$, $b$ and $c$
are simultaneously zero and that (\ref{moduli}) holds. See 
\cite{Hori:2004ja} for an explicit change of coordinates in a
different example.} Furthermore we note that special points on this
curve correspond to standard permutation branes 
\cite{Recknagel:2002qq}: for example for $a\neq 0$ and $b=0$ we may
use the projective equivalence to set $a=1$. Then $c$ must be a fifth
root of $-1$, leading  precisely to a permutation factorisation of the
form discussed in \cite{Brunner:2005fv,Enger:2005jk}. This
identification is also in agreement with the analysis of
\cite{Ashok:2004zb,Brunner:2005fv} where it was shown that one of these matrix
factorisations carries indeed the charge of a D2-brane.

\FIGURE[r]{
\includegraphics[width=70mm]{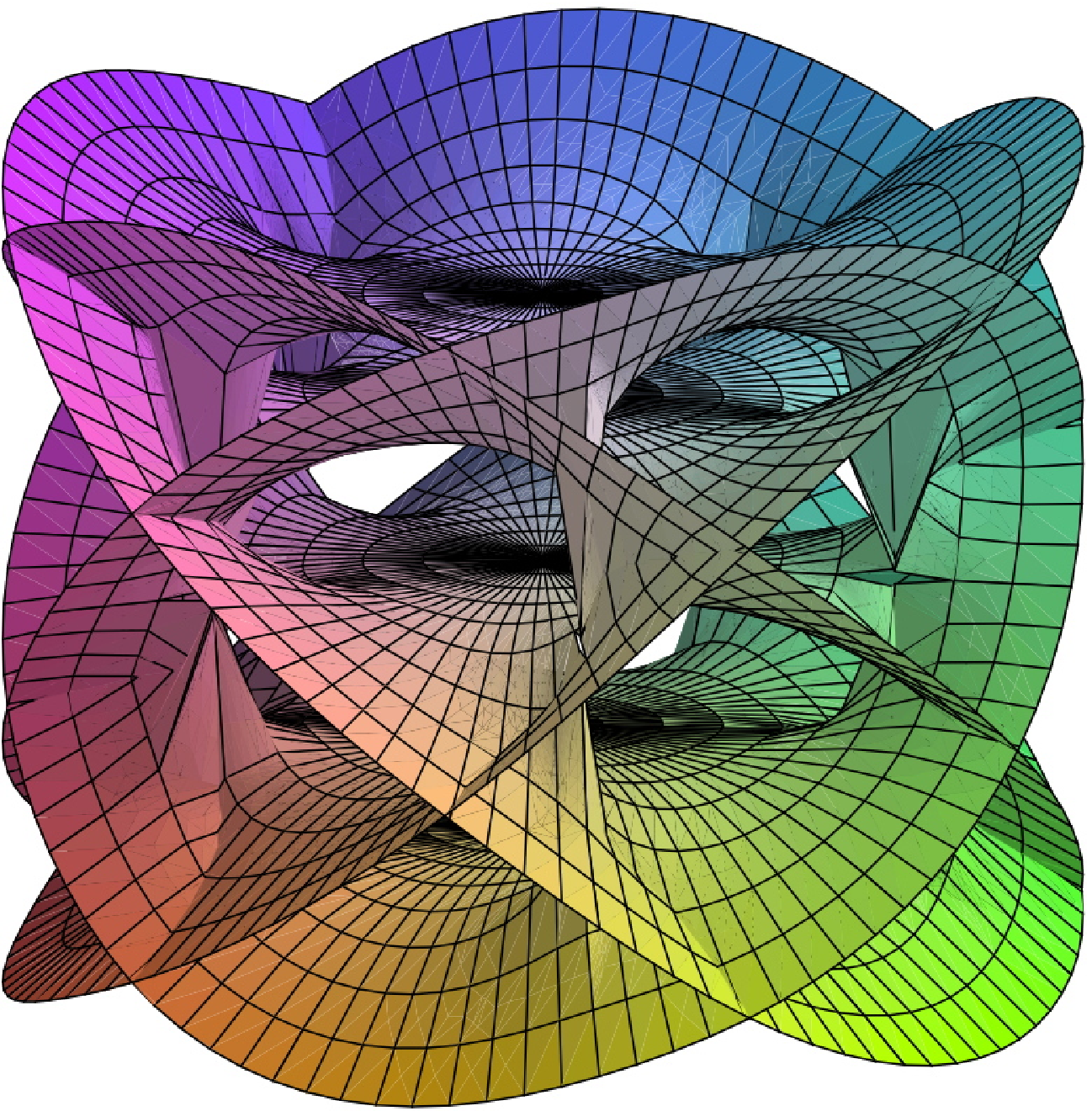}
\caption{Riemann surface associated to the D-brane moduli space,
consisting of five copies of the complex plane. The real and imaginary
part of $b$ have been plotted horizontally, the vertical axis is
the imaginary part of $c$. The five sheets arranged vertically at
$b=0$ reflect the five possibilities for $c^5=-1$.}   
}

\subsection{The fermionic spectrum}

The fact that these matrix factorisations form a 1-complex dimensional
moduli space means that at every point in the moduli space the open
string cohomology contains at least one fermion of $U(1)$-charge
one. Indeed, this is just the matrix factorisation analogue of the
fact that each such D-brane must have an exactly marginal boundary
operator in its spectrum. From a matrix factorisation point of view,
the corresponding fermion can be easily constructed. Since by 
assumption $a\neq 0$, we may always rescale the parameters so that $a=1$. 
Let us first consider a generic point in moduli space where 
$bc\neq 0$. We then have a family of factorisations parametrised by
$(b,c)$ subject to $b^5 + c^5 = -1$. As long as $c\neq 0$, we can
locally solve this equation for $c$, {\it i.e.}\ we can express
$c\equiv c(b)$, and thus obtain a matrix factorisation $Q(b)$. Since
$W_0$ does not depend on $(a,b,c)$, it then follows that   
\be
\{ Q(b), \partial_b Q(b) \} = 0 
\ee
which is precisely the condition for $\psi = \partial_b Q(b)$ to
define a fermion of the cohomology defined by $Q(b)$. For the case
under consideration, we find explicitly
\be\label{fermion1}
\psi_b \equiv \partial_b Q(b) = - x_3\, \pi^4 
- \frac{b^4}{c^4} x_3 \, \pi^5 
+ \left(\partial_b E_4\right)\, \bar\pi^4 
+ \left(\partial_b E_5\right)\,  \bar\pi^5 \ ,
\ee
where we have used that 
\be
\left. \frac{\partial b}{\partial c} \right|_a = - \frac{c^4}{b^4} \ . 
\ee
One can show by explicit computation (see appendix~A) that $\psi_b$ is
non-trivial in cohomology. Obviously, we could have equally expressed 
$b\equiv b(c)$ (for $b\neq 0$) and written 
$Q(a,b,c) \equiv Q(c)$. Then the derivation with respect to $c$ also
defines a fermion 
\be\label{fermion}
\psi_c \equiv 
\partial_c Q(c) = \frac{c^4}{b^4} x_3 \, \pi^4
+ x_3\, \pi^5 
+ \left(\partial_c E_4\right)\, \bar\pi^4 
+ \left(\partial_c E_5\right)\,  \bar\pi^5 \ .
\ee
It is easy to see that for $bc\neq 0$ so that both $\psi_b$ and
$\psi_c$ are well defined, $\psi_b\cong \psi_c$ in cohomology. In the
following we shall denote the equivalence class to which $\psi_b$ and
$\psi_c$ belong by $\psi_1$. More specifically, we shall usually
take $\psi_1\equiv \psi_b$ and assume that $c\neq 0$. 
\smallskip

\noindent
The full fermionic cohomology of $Q(a,b,c)$ at $U(1)$-charge
$1$ is however bigger: in addition to $\psi_1$ it also contains a
second fermion that we shall call $\psi_2$. This is explained in
appendix~A, where $\psi_2$ is explicitly constructed (for 
$c\neq 0$). In general, however, $\psi_2$ does not define a
modulus. In fact, using the Kapustin-Li formula \cite{Kapustin:2003ga}
one easily finds that  
\be\label{kap-li}
B_{\psi_2\psi_2\psi_2} = -\frac{2}{5}\eta^4 \frac{b^3}{c^9} \ . 
\ee
Unless $b=0$ the three-point function of $\psi_2$ on the boundary does
not vanish, and hence $\psi_2$ is not an exactly marginal boundary 
field \cite{Recknagel:1998ih}. This shows that at generic points 
in the moduli space (\ref{moduli}) there is only one exactly marginal 
operator, whereas at the special point $b=0$ an additional marginal 
operator appears, indicating an additional  branch of the moduli
space. This is in nice agreement with the geometric analysis of
section~1.1 since at $b=0$ the above moduli space intersects with the
branch where the roles of $J_1$ and $J_5$ can be interchanged.
In fact, this can also be seen from the explicit formula for $\psi_2$,
see (\ref{psi2ex}) in appendix~A.

The three-point function (\ref{kap-li}) verifies 
the superpotential term (\ref{bdlr}) that was already obtained in  
\cite{Ashok:2004xq} by other means. Furthermore, 
after rescaling $\psi_2 \mapsto \hat\psi_2 = c^3 \psi_2$, 
the $b$-dependence of the three-point function for $\hat\psi_2$ is
simply proportional to $b^3$. (Recall that $c\equiv c(b)$.) Using the
arguments of section 4.1 this then implies that, with respect to this
normalisation, the effective superpotential does not contain any 
higher order contributions (in $\psi_1$) to the term 
$\psi_1^3 \, \hat\psi_2^3$ in (\ref{bdlr}). This is in agreement with
the recent analysis of \cite{Aspinwall:2007cs}.

\section{Bulk perturbation }

Now we want to consider the bulk perturbation of the above Gepner
model by the bulk operator $\Phi$, {\it i.e.} we consider the
perturbed superpotential 
\be\label{bper}
W = W_0 + \lambda\, \Phi  \ , \qquad
\Phi = x_1^3 \, s^{(2)}(x_3,x_4,x_5) \ , 
\ee
where $s^{(2)}$ is the polynomial of section~1.1 that we expand 
as\footnote{Everything we are going to say is essentially unchanged if
we were to replace $x_1^3$ by an arbitrary third order polynomial in
$x_1$ and $x_2$.}  
\be\label{defo}
s^{(2)}(x_3,x_4,x_5) = \sum_{q+r+s=2} s^{(2)}_{qrs}\, \,
x_3^q \, x_4^r \, x_5^s \ .
\ee
{}From a conformal field theory point of view the perturbation is
generated by an exactly marginal bulk field in the $cc$ ring.
We want to understand what happens to the D-branes
described by the moduli space (\ref{moduli}) under this
perturbation. We shall be able to give a fairly complete description
of this problem by combining the ideas of \cite{Fredenhagen:2006dn}
with matrix factorisation techniques. In particular, this will allow
us to calculate the effective superpotential for the boundary
parameters $(a,b,c)$ exactly.   

One way to address this problem is to study the deformation theory of
matrix factorisations, following \cite{Hori:2004ja} (see also 
\cite{Ashok:2004xq}). Suppose that $Q_0$ is a factorisation of
$W_0$. Then we ask whether we can find a deformation $Q$ of $Q_0$,
{\it i.e.}
\be
Q= Q_0 + \lambda Q_{1} + \lambda^2 Q_{2} + \cdots
\ee
such that $Q^2 = W_0 + \lambda \Phi$. Expanding this equation to first
order in $\lambda$, we find the necessary condition that $\Phi$ must be
exact with respect to $Q_0$, {\it i.e.}\  of the form 
$\Phi = \{ Q_0, \chi \}$ for some $\chi$. In general this condition
will not be met; for example for the case at hand where 
$Q_0\equiv Q(a,b,c)$ and $\Phi$ is given by (\ref{bper}), we find that
$\Phi$ is exact if and only if 
\be\label{constraint}
a^5 + b^5 + c^5 = 0 \qquad \hbox{and} \qquad
s^{(2)}(a,b,c)=0 \ .
\ee
On the other hand, if this condition {\it is met}, it is easy to see
that we {\it can} in fact extend the matrix factorisation for
arbitrary (finite) values of $\lambda$. Indeed, if we consider the
same ansatz as in (\ref{ansatz}), it is clear 
that we can find a joint solution to $J_1=J_4=J_5=0$ and
$W=W_0+\lambda \Phi=0$ if $(a,b,c)$ satisfies (\ref{constraint}). It
then follows by the same arguments as above that there exists a
matrix factorisation for all values of $\lambda$ (that is by
construction a deformation of $Q(a,b,c)$). 

Unless $s^{(2)}\equiv 0$, the set of constraints (\ref{constraint})
has only finitely many discrete solutions; in fact, counting
multiplicities, there are precisely $10$ solutions, as follows from 
Bezout's theorem. This ties in nicely with our geometric expectations
since at a generic point in the complex structure moduli space only
finitely many holomorphic 2-cycles exist.

\subsection{Combining with conformal field theory}

As we have just seen, for $\lambda=0$ we have a one-parameter family of
superconformal D2-branes, while for $\lambda\neq 0$ only discrete
possibilities remain. The situation is therefore very analogeous to
the example studied in  \cite{Fredenhagen:2006dn}. There a general
conformal field theory analysis of this problem was suggested that we
now want to apply to the case at hand.

In \cite{Fredenhagen:2006dn} the coupled bulk and boundary
deformations of a boundary conformal field theory were studied, and
the resulting renormalisation group identities were derived. It
was found that an exactly marginal bulk operator may cease to be
exactly marginal in the presence of a boundary. If this is the case it
will induce a renormalisation group flow on the boundary that will
drive the boundary condition to one that is again conformal with
respect to the deformed bulk theory. If we denote the boundary
coupling constant corresponding to the boundary field $\psi_j$ 
of conformal weight $h_j$ by $\mu_j$, then the perturbation by the
exactly marginal bulk operator $\lambda \Phi$ will induce the RG
equation  
\be\label{RG}
\dot\mu_j = (1- h _j)\mu_j +  \frac{\lambda}{2} B_{\Phi \psi_j} 
+ {\cal O}(\mu\lambda, \lambda^2, \mu^2)\ , 
\ee
where $B_{\Phi \psi_j}$ is the bulk-boundary operator product
coefficient. Since the first term in (\ref{RG}) damps the flow of any
irrelevant operators, it is sufficient to study this equation only for
the marginal or relevant boundary fields, {\it i.e.} for those that
satisfy $h_j\leq 1$. 

For the case at hand, we do not have an explicit conformal field
theory description of the D-branes away from the specific points where
$abc=0$. On the other hand, we know (based on supersymmetry) that the
open string spectrum will not contain any relevant (tachyonic)
operators. Furthermore, the above discussion suggests that
everywhere in moduli space each brane has precisely two marginal 
operators in its spectrum, namely the operators corresponding to the
open string fermions described by $\psi_1$ and $\psi_2$ --- see
appendix~A for details. The two boundary operators $\psi_1$ and
$\psi_2$ are topological, and so is the bulk perturbation $\Phi$. In
particular, this implies that we can determine the coefficients
$B_{\Phi \psi_1}$ and $B_{\Phi \psi_2}$ that are important for the RG
equations using {\it topological methods}, without having to solve the 
full conformal field theory (which would be impossibly difficult)! 

Using the Kapustin-Li formula (\ref{KL}) we find (we are
working in a patch where $a=1$) 
\be\label{psi2}
B_{\Phi\psi_2} = 0 
\ee
for all $(a,b,c)$, as well as 
\be
B_{\Phi \psi_b} = 
	\frac{\eta^4}{25} c^{-4} s^{(2)}(1,b,c) \ ,
\ee
and similarly for 
\be
B_{\Phi \psi_c} =  
	- \frac{\eta^4}{25} b^{-4} s^{(2)}(1,b,c) \ .
\ee
All of these calculations were performed in the unperturbed bulk
theory. Since the bulk-boundary coupling between $\Phi$ and $\psi_2$
vanishes (\ref{psi2}), this field is not switched on by $\Phi$. The RG
flow  will therefore only involve $\psi_1$, and for this we find
\be\label{bdot}
\dot{b} = \lambda\frac{\eta^4}{50} c^{-4} s^{(2)}(1,b,c) \ ,
\ee
or 
\be\label{cdot}
\dot{c} = - \lambda\frac{\eta^4}{50} b^{-4} s^{(2)}(1,b,c) \ .
\ee
In particular, we see that the solutions to (\ref{constraint}) are
precisely the fixed points under the RG equation. Thus any brane
described by $(a,b,c)$, will flow to one of these $10$ fixed points
under the RG flow.

\subsection{RG flow as gradient flow}

Actually, the above RG flow is a gradient flow, as was also the case
in the example studied in \cite{Fredenhagen:2006dn}.\footnote{For
exactly marginal bulk deformations this may in fact follow from the
analysis of \cite{Friedan:2003yc}.} In fact, we can integrate the RG
equation for $b$ in (\ref{bdot}) to   
\be\label{Wb}
\dot{b} = \partial_b {\cal W}(a,b,c) \ ,
\ee
where ${\cal W}(a,b,c)$ is evaluated on the moduli space
(\ref{moduli}) with  $a^5+b^5+c^5=0$ and we have rescaled
$a=1$. Similarly, the same function ${\cal W}(a,b,c)$ also controls
the RG equation for $c$ in (\ref{cdot}) 
\be\label{Wc}
\dot{c} = \partial_c {\cal W}(a,b,c) \ , 
\ee
where again $a=1$ and we regard $b$ as a function of $c$ via the
constraint $a^5+b^5+c^5=0$. To determine ${\cal W}(a,b,c)$ explicitly
we need to integrate
\be
\int_{b_0}^b db' B_{\Phi \psi_b} = \frac{\eta^5}{25} 
\int_{b_0}^b db' c^{-4} s^{(2)}(a,b',c) \ .
\ee
The integral is along a line on the Riemann surface starting at a fixed
reference point $b_0$ that we take to be $0$ and ending at $b$. Since
$b$ parametrises the brane moduli space, it has a natural physical
interpretation as the position of the brane.  The integrand is a
holomorphic one-form on the Riemann surface parametrising the moduli
space, see appendix~B for more details. The potential therefore has a
natural geometric interpretation as the Abel-Jacobi map associated 
to a one-form on the Riemann surface whose points label the brane
positions. Which particular one-form is to be integrated is determined
by the bulk deformation under consideration. 

Since the integrals of such forms are known, we can give explicit
formulae for ${\cal W}(a,b,c)$ in each patch. As explained in
appendix~B, in the patch where $a=1$ and $c\neq 0$ (so that
$c\equiv c(b)$ is well defined) one obtains
\be\label{superex}
{\cal W}(1,b,c)
= \lambda\frac{\eta^4}{50}\sum_{q+r+s=2}\frac{1}{r+1}s^{(2)}_{qrs}
  (-b)^{r+1} \,
  _2{\rm F}_1(\tfrac{r+1}{5},1-\tfrac{s+1}{5};1+\tfrac{r+1}{5};-b^5)
\ . 
\ee
It is also checked there that this function satisfies both (\ref{Wb})
and (\ref{Wc}). 

By combining abstract conformal field theory arguments with
topological methods we can thus give a complete description of the RG
flow: the D2-brane simply follows the gradient flow of ${\cal W}$ to
arrive at one of its local minima, which are precisely the points
characterised by (\ref{constraint}). As in \cite{Fredenhagen:2006dn},
in the RG scheme in which we always remain in the original moduli
space, this analysis is exact in the boundary moduli, and first order
in the bulk coupling constant. Obviously the picture we have found
ties in very nicely with the geometric expectations of section~1.1.

We should note that it is crucial in this analysis that the bulk
perturbation by $\Phi$ does not switch on $\psi_2$, {\it i.e.}\ that 
$B_{\Phi \psi_2}=0$. Otherwise the bulk perturbation would switch on a
boundary field that would lead us out of the original moduli
space and we would not be able to iterate the RG equations. This is
the reason why we restricted our analysis to the bulk perturbations of
the form described in (\ref{defo}).

\section{Superpotentials}

The function ${\cal W}$ has actually an interpretation in terms of the
effective spacetime superpotential, as we shall now explain. 

\subsection{Gradient flow of the superpotential}

In the above we have seen explicitly that the RG flow is a gradient  
flow of a potential. This potential is precisely the contribution to
the effective superpotential ${\cal W}$ that is first order in the
bulk field  $\Phi$ and exact in the boundary field $\psi_1$. To see
this we simply note that the term that appears on the right hand side
of (\ref{RG}) is the bulk-boundary coefficient that involves one
insertion of the bulk field $\Phi$ and one insertion of the boundary
field $\psi_1$ (that couples to $\mu$).  This bulk-boundary correlator
was evaluated at an {\it arbitrary} point in the brane moduli space;
if we start around any given point of the brane moduli space, the
above expression  therefore involves an arbitrary number of insertions
of $\psi_1$ (that allow one to move around this brane moduli
space). Thus the right-hand-side of (\ref{RG}) is the generating
function describing symmetrised correlators involving an arbitrary
number of boundary fields $\psi_1$, together with one insertion of the
boundary field $\psi_1$ and one insertion of the bulk field $\Phi$. We
can produce the insertion of the boundary field $\psi_1$ by taking a 
derivative with respect to the corresponding boundary coupling
constant. It thus follows that the function ${\cal W}$ (that we
obtained by integrating up the right hand side of (\ref{RG}))
is precisely the generating function of one bulk field $\Phi$ with an
arbitrary number of boundary fields. It therefore defines the
corresponding contribution of the effective superpotential.

It is also clear from this argument that this method can be applied to
calculate the corresponding terms of the effective superpotential for 
an arbitrary bulk deformation, not just one of the form
(\ref{defo}). For the other cases, the result is however trivial: the
complex structure deformations (\ref{defo}) are the only monomials 
(instead of $x_1^3$ we may also allow for an arbitrary third order
polynomial in $x_1$ and $x_2$) for
which the bulk-boundary OPE coefficient with $\psi_1$ is
non-zero. Thus to first order in the bulk perturbation the
above terms are the only terms that appear in the effective
superpotential. It should also be obvious how to perform the same
analysis for the other ($45$) families of D2-branes.   

\subsection{Interpretation in terms of Chern-Simons theory}

The observation that superpotentials can be viewed as Abel-Jacobi
maps for certain one-forms on Riemann surfaces labelling the brane 
positions have appeared before in
\cite{Aganagic:2000gs,Aganagic:2001nx} (see also \cite{Clemens}).   
In their approach, these superpotentials were calculated using
geometric methods, which yields the exact result on the
B-side. Namely, the superpotential is given by the action of a
holomorphic Chern Simons theory living on the brane
\cite{Witten:1992fb} which for branes wrapping holomorphic curves in
the internal dimensions can be reduced to an integral of the
holomorphic 3-form over a 3-chain. These integrals can be explicitly
evaluated in the examples studied in
\cite{Aganagic:2000gs,Aganagic:2001nx} and are effectively reduced to 
one-dimensional integrals along a path with two end-points on the
Riemann surface. These one-dimensional integrals give the
superpotential in terms of the Abel Jacobi map for the one-form,
similar to our case.  

In later developments,
\cite{Mayr:2001xk,Lerche:2001cw,Lerche:2002yw,Govindarajan:2001zk} the
3-chain integrals mentioned above appeared in certain differential
equations extending the well-known Picard Fuchs equations governing
the closed string sector. In the context of the quintic, such a
differential equation has been proposed for a different D-brane in
\cite{Walcher:2006rs}, resulting in a full instanton expansion on the
mirror A-side (see \cite{Brunner:2004mt,Govindarajan:2006uy} for a
result on the torus). In our example, the appearance of
hypergeometric functions as effective potentials suggests to consider
the hypergeometric differential equation of which these are
solutions. We speculate that this differential equation is part of the
Picard Fuchs equations governing the open string sector in the sense
of \cite{Mayr:2001xk,Lerche:2001cw,Lerche:2002yw}.

\subsection{Integrability}

The fact that the bulk-boundary correlator could be integrated
up to a generating function of the boundary fields $\psi_a$ with the
insertion of the given bulk field $\Phi$ is something we can show in 
generality. If we denote the relevant bulk-boundary correlator by
$B_{\Phi \psi_a}$, then the integrability condition is simply that  
\be\label{sym}
\partial_{a_1} B_{\Phi \psi_{a_2}} = 
\partial_{a_2} B_{\Phi \psi_{a_1}} \ .
\ee
Similar relations are well-known to hold in the bulk \cite{DVV}; 
for boundary correlators there are usually ordering ambiguities
for the operators on the boundary, and hence similar relations can only
be expected to hold for the symmetrised correlation functions.
However, the disc correlation functions are still cyclically
symmetric in the boundary insertions and totally symmetric
in the bulk insertions \cite{HLL} from which one can infer that
(\ref{sym}) holds. The following is essentially a review of the
argument of \cite{HLL}, which again is similar to \cite{DVV}.
We will work in the topologically twisted theory, where $G^+$ becomes  
the BRST current $Q$, while $G^-$ becomes an $h=2$ field of the
topological theory. The starting point are the Ward identities for the 
open-closed correlators \cite{HLL}
\ba
0=&& \langle \oint \xi(w) G(w) \psi_{a_1} (\tau_1) \cdots 
\psi_{a_m} (\tau_m)
\Phi_{i_1}(z_1, \bar{z}_1) \cdots \Phi_{i_s}(z_s, \bar{z}_s) \rangle 
\nonumber \\
 =&&
\sum_{k=1}^m \pm \xi(\tau_k) \,
\langle \psi_{a_1} (\tau_1) \cdots \psi_{a_k}^{(1)}(\tau_k) \cdots 
\psi_{a_m} (\tau_m)  \, 
\Phi_{i_1}(z_1, \bar{z}_1) \cdots \Phi_{i_s}(z_s, \bar{z}_s) \rangle 
\nonumber \\
&& \quad \pm \sum_{l=1}^s \xi (z_l) \, 
\langle \psi_{a_1} (\tau_1) \cdots \psi_{a_m} (\tau_m) \, 
\Phi_{i_1}(z_1,\bar{z}_1) \cdots 
\Phi_{i_l}^{(1,0)}(z_{l},\bar{z}_{l}) \cdots 
\Phi_{i_s}(z_s,\bar{z}_s) \rangle
\nonumber \\
&& \quad \pm \sum_{l=1}^s \bar\xi (\bar{z}_l) \,
\langle \psi_{a_1} (\tau_1) \cdots \psi_{a_m} (\tau_m) \, 
\Phi_{i_1}(z_1,\bar{z}_1) \cdots \Phi_{i_l}^{(0,1)}(z_l,\bar{z}_l) 
\cdots  \Phi_{i_s}(z_s,\bar{z}_s) \rangle.\qquad
\ea
Here, $\psi_{a_j}$ denote the boundary fields at points $\tau_j$
on the  real line, and $\Phi_i$ are the bulk insertions. Furthermore,
we have  
\be
\vert \psi^{(1)}_a\rangle = G_{-1} \vert \psi_a \rangle \ ,
\ee 
with a similar definition for $\Phi^{(1,0)}$ and
$\Phi^{(0,1)}$. Finally, the contour integral is taken to surround all
fields, and $\xi$ is a globally defined holomorphic
vector field on the upper half plane
\be
\xi (w) = aw^2 + bw +c\ ,
\ee
where $a,b,c$ are real constants. The above signs are uniquely
determined by the statistics of the various fields; we shall be more
specific momentarily.

We are interested in the special case of a correlator with one bulk
insertion $\Phi$, and two boundary insertions $\psi_l$ and
$\psi_m$. The vector field $\xi$ can then be chosen such that the term
with an integrated bulk insertion drops out from the Ward identity 
\be
\xi(w) = (w-z)(w-\bar{z}) \ .
\ee
The Ward identities then give
\be \label{ward}
\xi(\tau_l) \; \langle \phi(z)\, \bar{\phi} (\bar{z}) \, 
\psi_l^{(1)}(\tau_l) \, \psi_m(\tau_m) \rangle =
\xi(\tau_m) \; \langle \phi(z) \, \bar{\phi} (\bar{z}) \,
\psi_l(\tau_l) \, \psi_m ^{(1)}(\tau_m) \rangle \ . 
\ee
Here we have used the usual doubling trick to represent the bulk field
$\Phi(z,\bar{z})$ in terms of a chiral field $\phi(z)$ on the upper
half plane, together with its image  $\bar\phi(\bar{z})$ on the lower
half plane. We have furthermore used that the statistics of the 
boundary insertions that are of interest to us is fermionic. One way
to see this is in terms of the Landau-Ginzburg theory, where
perturbations of the BRST operator $Q\to Q+\lambda \psi$ always
involve boundary fermions. 

Following \cite{DVV} we now use the conformal symmetry to evaluate
these four-point functions further. For the function on the left hand
side we consider the M\"obius transformation
\be
\hat{f}(u) = \frac{u-\bar{z}}{u-z} \, 
                \frac{\tau_m - z}{\tau_m -  \bar{z}} 
\ee
which maps the points $z, \bar{z},\tau_m$ and $\tau_l$ to 
$\infty, 0, 1$ and 
\be
\hat\zeta = \frac{\tau_l-\bar{z}}{\tau_l-z} \, 
                \frac{\tau_m - z}{\tau_m -  \bar{z}} \ . 
\ee
In the topologically twisted theory $\phi$ and $\bar\phi$ have
conformal weight zero, as has $\psi_a$, while $\psi_a^{(1)}$ has
conformal weight one. It thus follows from the usual conformal
transformation properties that 
\begin{eqnarray}
\langle \phi(z) \, \bar{\phi} (\bar{z}) \, 
\psi_l^{(1)}(\tau_l) \, \psi_m(\tau_m) \rangle & = & 
\hat{f}{}'(\tau_l) \, 
\langle \phi(\infty)\, \bar{\phi} (0) \, 
\psi_l^{(1)}(\hat\zeta) \, \psi_m(1) \rangle \nonumber \\
& = & 
\frac{\bar{z}-z}{(\tau_l-z)^2} \, 
                \frac{\tau_m - z}{\tau_m -  \bar{z}} \,
\langle \phi(\infty) \, \bar{\phi} (0) \, 
\psi_l^{(1)}(\hat\zeta) \, \psi_m(1) \rangle \ . \label{eq1}
\end{eqnarray}
Similarly, we use the M\"obius transformation 
\be
f(u) = \frac{u-\bar{z}}{u-z} \, 
                \frac{\tau_l - z}{\tau_l -  \bar{z}} 
\ee
to rewrite the right hand side as 
\be
\langle \phi(z) \, \bar{\phi} (\bar{z}) \,
\psi_l(\tau_l) \, \psi_m ^{(1)}(\tau_m) \rangle = 
\frac{\bar{z}-z}{(\tau_m-z)^2} \, 
                \frac{\tau_l - z}{\tau_l -  \bar{z}}  \,
\langle \phi(\infty) \, \bar{\phi} (0) \, 
\psi_l(1) \, \psi_m^{(1)}(\zeta) \rangle \ ,
\ee
where $\zeta = \hat\zeta^{-1}$. Using the explicit form of 
$\xi(\tau_l)$ and $\xi(\tau_m)$ it then follows from the Ward identity
(\ref{ward}) that 
\be
\langle \phi(\infty) \, \bar{\phi} (0) \, 
\psi_l^{(1)}(\hat\zeta) \, \psi_m(1) \rangle  = 
\zeta^2 \, 
\langle \phi(\infty) \, \bar{\phi} (0) \, 
\psi_l(1) \, \psi_m^{(1)}(\zeta) \rangle \ . 
\ee
The right hand side of (\ref{sym}) is now the integral 
\begin{eqnarray}
\partial_{l} B_{\Phi \psi_m} & = &  \int_{C_+} d\hat\zeta \; 
\langle \phi(\infty) \, \bar{\phi} (0) \, 
\psi_l^{(1)}(\hat\zeta) \, \psi_m(1) \rangle \nonumber \\
& = & \int_{C_+} d\hat\zeta \, \hat\zeta^{-2}\;
\langle \phi(\infty) \, \bar{\phi} (0) \, 
\psi_l(1) \, \psi_m^{(1)}(\zeta) \rangle \nonumber \\
& = & - \int_{C_-} d\zeta \;
\langle \phi(\infty) \, \bar{\phi} (0) \, 
\psi_l(1) \, \psi_m^{(1)}(\zeta) \rangle \nonumber \\
& = & + \int_{C_+} d\zeta \;
\langle \phi(\infty) \, \bar{\phi} (0) \, 
\psi_l(1) \, \psi_m^{(1)}(\zeta) \rangle \nonumber \\
& = & \partial_{m} B_{\Phi \psi_l} \ . 
\end{eqnarray}
Here, the contour $C_+$ denotes the unit circle with positive 
orientation and $C_-$ is the circle with opposite orientation. This
proves the desired integrability condition.

\section{Conclusions}

In this paper we have studied the behaviour of a class of D2-branes on
the quintic under complex structure deformations of the closed string
background. At the Gepner point there are $50$ one-dimensional
moduli spaces of D2-branes intersecting over certain curves. As one
switches on a complex structure deformation of the Gepner point, only
a finite number of the corresponding $\IP_1$'s remain \cite{AK}. This
implies that there is an effective space-time superpotential, that
possesses a flat direction at the Gepner point, while it only has 
finitely many discrete minima at generic values of the bulk
parameters. From a world-sheet point of view, the interpretation is
that the bulk deformation is not truly marginal on world-sheets with
boundaries. Rather, it induces a non-trivial RG flow at the boundary,
that drives the boundary condition to one that is compatible with the
changed background geometry. The relevant driving term can be
calculated using matrix factorisation techniques, and the RG flow
turns out to be a gradient flow of a potential. This potential can be
identified with (a certain contribution to) the effective space-time
superpotential that we can calculate explicitly. The calculation is
first order in the bulk perturbation, but exact to all orders in the
boundary couplings.   

We have only studied a certain family of branes on the quintic; it
should however be straightforward to generalise our methods to other
Calabi-Yau manifolds and other classes of branes. As we have indicated
in the main part of the paper one may expect that the integrated bulk-boundary
couplings generalise the familiar bulk periods, and hence the superpotential 
terms
obey a Picard-Fuchs type equation \cite{Mayr:2001xk,Lerche:2001cw,Lerche:2002yw}; it would
be very interesting to analyse this in detail. Our result should 
also have an interesting interpretation in the context of mirror
symmetry. We have calculated the superpotential for a B-type D-brane
on a compact Calabi-Yau manifold which, in particular, is exact to all
orders in $\alpha'$. On the mirror A-side, the superpotential vanishes
perturbatively and can only be generated non-perturbatively via disc
instantons  \cite{Kachru:2000ih,Kachru:2000an}. Our expression for the
B-type superpotential on the D2-branes should therefore have an 
interpretation in terms of disc instantons on the mirror A-side. One
would expect that a Picard-Fuchs like equation would be helpful in
analysing this question \cite{Lerche:2002yw}. We hope to return to
these issues in the near future.

\appendix

\section{The cohomology of the factorisations}

In this appendix we want to determine the full fermionic cohomology of
$U(1)$-charge $1$ for the factorisations $Q(a,b,c)$ (\ref{ansatz})
with $a\neq 0$. First we observe that the coordinates
involved in $J_1$ and $E_1$ (namely $x_1$ and $x_2$) do not appear in
$J_4, E_4$ or $J_5, E_5$. Therefore the cohomology $H$ of $Q$
separates into 
\be
H(Q) = H(Q_1)\odot H(Q_2)\ ,
\ee  
where $Q_1$ and $Q_2$ are the separate factorisations
\ba
Q_1 &=& \pi^1 J_1+\bar\pi^1 E_1\ ,\nonumber\\
Q_2 &=& \pi^4 J_4 + \pi^5 J_5 + \bar\pi^4 E_4 +\bar\pi^5 E_5 \ . 
\ea

\noindent The explicit polynomials are
\ba
J_1 &= x_1-\eta x_2 \qquad E_1 &
       = \prod_{\eta'^5=-1,\eta'\ne\eta}(x_1-\eta'x_2) \nonumber\\
J_4 &= ax_4-b x_3 \qquad E_4 &= \hphantom{-}\frac{1}{a^5}\left(
	b^4x_3^4 + ab^3x_3^3x_4 + a^2b^2x_3^2x_4^2
        +a^3bx_3x_4^3+a^4x_4^4\right) \\
J_5 &= cx_3-ax_5 \qquad E_5 &=-\frac{1}{a^5}\left(
	c^4x_3^4+ac^3x_3^3x_5+a^2c^2x_3^2x_5^2
        +a^3cx_3x_5^3+a^4x_5^4\right)\ . \nonumber
\ea
The cohomology of $Q_1$ has been calculated in
\cite{Ashok:2004zb,Brunner:2005fv,Enger:2005jk}, and consists of four  
bosonic elements of $U(1)$-charge $0$, $\tfrac{2}{5}$,
$\tfrac{4}{5}$ and $\tfrac{6}{5}$, respectively; it does not
contain any fermions at all. Thus in order to obtain a fermion of the 
full factorisation, we need to tensor one of these bosons with a
fermion from $Q_2$. We are only interested in fermions of $Q$ of total 
$U(1)$-charge $1$. Since the $U(1)$-charge of the fermions in $Q_2$ is
always positive, there are three cases to consider:
the fermions in the cohomology of $Q_2$ can have $U(1)$-charges $1$,
$\frac{3}{5}$ or $\frac{1}{5}$ which together with the boson of $Q_1$
of $U(1)$-charges $0$, $\frac{2}{5}$ or $\frac{4}{5}$, respectively,
then produce a fermion of total $U(1)$-charge $1$. Thus it is
sufficient to analyse the fermionic cohomology of $Q_2$ for these
three $U(1)$-charges separately.  

\subsection{The $Q_2$-fermions of charge $1$}

The general $Q_2$-closed fermion has an expansion (the closure
conditions force the absence of any higher powers of boundary
fermions)  
\be
\psi=\pi^4 p_4 + \bar\pi^4 m_4 + \pi^5 p_5 + \bar\pi^5 m_5  \ ,
\ee
where we have dropped some exact terms --- see (\ref{exact}) below. 
The requirement that $\psi$ has $U(1)$-charge $1$ implies that 
$p_4$ and $p_5$ are polynomials of degree $1$ (thus each $p_i$ has $3$
parameters)  while $m_4$ and $m_5$ are polynomials of degree $4$ (with
$15$ parameters each), giving in total $36$ parameters. The condition
that $\psi$ is closed implies further that 
\be
J_4 \, m_4 + J_5 \, m_5 + E_4 \, p_4 + E_5 \, p_5 = 0 \ .
\ee
The left hand side is a homogeneous polynomial of degree $5$, and hence
represents $21$ conditions. We have checked (using standard matrix
techniques) that these $21$ conditions are independent. This implies
that the space of closed fermions of the $U(1)$-charge $1$ is 
$15$-dimensional. 

It remains to determine how many of them are exact. To see this we
make the following ansatz for the most general boson, 
\ba
\Lambda = &\hat a + \hat b \pi^4\bar\pi^4 + \hat c\pi^4\pi^5 
		+ \hat d\pi^4\bar\pi^5  
		+ \hat e\bar\pi^4\pi^5
		 +\hat f\bar\pi^4\bar\pi^5 
		+ \hat g\pi^5\bar\pi^5 
		+ \hat h\pi^4\bar\pi^4\pi^5\bar\pi^5 \ . 
\ea
Then
\begin{align}
[Q,\Lambda]=
&\pi^4\left(-\hat bJ_4-\hat dJ_5-\hat cE_5\right) 
+ \pi^5\left(\hat eJ_4-\hat gJ_5+\hat cE_4\right) \nonumber \\
&+\bar\pi^4\left(\hat bE_4-\hat eE_5-\hat fJ_5\right) 
	+ \bar\pi^4\left(\hat dE_4+\hat gE_5+\hat eJ_4\right) 
\nonumber \\
&-\pi^4\bar\pi^4\pi^5 \hat hJ_5 + \pi^4\bar\pi^4\bar\pi^5 \hat hE_5 
-\pi^4\pi^5\bar\pi^5 \hat hJ_4 + \bar\pi^4\pi^5\bar\pi^5 \hat hE_4 
\ . \label{exact}
\end{align}
Consistency with the ansatz for $\psi$ requires $\hat h=0$ and 
$\hat c=0$. Moreover $\hat a$ can be set to zero, too. The other
parameters must be polynomials of degree $0$, except for $\hat f$
which has to have degree $3$ (and therefore $10$ parameters). In total 
the space of exact fermionis is described by $14$ parameters. Again,
using standard matrix methods, we have shown that these $14$
parameters are linearly independent. This implies that 
the fermionic cohomology of $Q_2$ of $U(1)$-charge $1$ is
$1$-dimensional. A representative of the corresponding cohomology
class for $Q$ is (for $c\neq 0$)  
\be
	\psi_1 = \partial_b Q 
\ee
or explicitly
\ba
\psi_1 = &-&x_3 \pi^4 
	+\frac{1}{a^5}\left[
	4b^3x_3^4 + 3ab^2x_3^3x_4 
        + 2a^2bx_3^2x_4^2+a^3x_3x_4^3\right]\bar\pi^4 \\
	&-& \frac{b^4}{c^4} x_3 \pi^5
		+\frac{b^4}{a^5c^4}\left[
	4c^3 x_3^4+3ac^2x_3^3x_5 
		+ 2a^2cx_3^2x_5^2+a^3x_3x_5^3\right]\bar\pi^5 \ .
\ea

\subsection{The $Q_2$-fermions of charge $\frac{3}{5}$}

The same arguments can be used to determine the fermions of $U(1)$-charge
$\tfrac{3}{5}$. In this case, $p_4$ and $p_5$ have both degree $0$ 
({\it i.e.}\ are constants) while $m_4$ and $m_5$ have both degree
$3$ (with 10 parameters each), giving rise to $22$ parameters.
The closure condition is now given by a polynomial of degree $4$,
leading to $15$ (independent) equations. Thus the space of closed
fermions is in this case $7$-dimensional.

For exact fermions we find that they are described by bosons $\Lambda$
with $\hat a=0$, $\hat b=0$, $\hat d=0$, $\hat c=0$, $\hat e=0$, 
$\hat g=0$, $\hat h=0$ and $\hat f$ a polynomial of degree $2$
(with $6$ parameters). Thus there are $6$ different exact fermions,
and we have checked that they are in fact linearly independent. This
implies that there is precisely one fermion of charge $\tfrac{3}{5}$
in the cohomology of $Q_2$. A representative of the corresponding
cohomology class for $Q$ is given by (for $c\neq 0$)  
\begin{align}
\psi_2 = x_1\, \partial_b \, \Bigl[
b\pi^4 - c\pi^5
- (b^4x_3^3+b^3x_3^2x_4+b^2x_3x_4^2+bx_4^3)&\bar\pi^4 \nonumber \\
+ (c^4x_3^3 + c^3x_3^2x_5 + c^2x_3x_5^2 + cx_5^3)&\bar\pi^5 
		\Bigr] \ , 
\end{align}
or, since $\psi_1$ is proportional to $x_3$, 
\be\label{psi2ex}
\psi_2 = \frac{x_1}{x_3} \psi_1 \ .
\ee

\subsection{The $Q_2$-fermions of charge $\frac{1}{5}$}

For fermions of charge $\frac{1}{5}$, our ansatz has $12$ parameters,
and the closure condition leads to $9$ linearly independent
conditions. Thus there are $3$ different closed fermions. In
$\Lambda$, all parameters are zero except $\hat f$, which is a
polynomial of degree $1$ with $3$ independent parameters. This implies
that all $3$ closed fermions are in fact exact, and hence that the
cohomology is trivial.

\section{Calculating the superpotential}

In this appendix we give details about how to calculate the effective
superpotential ${\cal W}$ explicitly. We begin by recalling some
facts about differentials on Fermat curves.

\subsection{Differentials on the Fermat curve and their integrals} 

Let us consider the Fermat curve defined by 
\be\label{fermat}
	\hat{b}^5+\hat{c}^5=1 \ .
\ee
For $a=1$ this is the curve that describes the brane moduli space 
$1+b^5+c^5=0$ provided we identify $\hat{b}=-b$ and $\hat{c}=-c$. The
general theory of globally defined differentials is described in
\cite{Lang}. The simplest class of differentials, the differentials of
the first kind, are those that are holomorphic on the full curve. They
are of the form
\be\label{first}
\omega_{rs} = \hb^r \hc^s\,\frac{\tfrac{1}{5}d(\hb^5)}{\hb^5 \hc^5} 
            = \hb^{r-1} \hc^{s-1}\frac{d\hb}{\hc^4} \ ,
\ee
where $r,s,\geq 1$. Since $\hb^4 d\hb =-\hc^4 d\hc$ this is equivalent
to  
\be\label{second}
\omega_{rs} = -\hb^{r-1} \hc^{s-1}\frac{d\hc}{\hb^4} \ .
\ee
The first formula (\ref{first}) is defined on the patch of the moduli
space where $\hc\neq 0$, while the second (\ref{second}) is defined
for $\hb\neq 0$. Since on (\ref{fermat}) $\hb \hc\neq 0$ at least one
of these two expressions is everywhere well-defined. In particular,
this therefore proves that the differentials $\omega_{r,s}$ are
holomorphic for finite $\hb$ and $\hc$. The only potential poles may
thus appear at $\hb,\hc=\infty$. Expanding around $\hb=\infty$ shows
that the differentials are finite as long as $r+s\le 4$. Therefore we
find the holomorphic differentials (for $\hc\neq 0$)
\be
	\frac{1}{\hc^4}d\hb,\;
	\frac{1}{\hc^3}d\hb,\;
	\frac{1}{\hc^2}d\hb,\;
	\frac{b}{\hc^4}d\hb,\;
	\frac{\hb^2}{\hc^4}d\hb,\;
	\frac{\hb}{\hc^3}d\hb \ .
\ee
In fact this is a basis for the holomorphic differentials on the
curve. Its number is equal to the genus of the curve.

\subsubsection{Integrating the holomorphic differentials}

In order to calculate the effective superpotential we need to
integrate these holomorphic differentials. For all of them the answer
can be expressed in terms of a hypergeometric function. In fact in the
chart where $\hc\neq 0$ we have 
\be
\label{EQHypergeoInt1}
\int_0^{\hb} \omega_{rs}
=\int_0^{\hb} d \tilde{b}\,\frac{\tilde{b}^{r-1}\hc(\tilde{b})^{s-1}}{
\hc(\tilde{b})^4} 
= \frac{1}{r}\, \hb^r\,
{}_2{\rm F}_1(\tfrac{r}{5},1-\tfrac{s}{5};1+\tfrac{r}{5}; \hb^5) \ .
\ee
On the other hand in the chart with $\hb\neq 0$ we get instead 
\be
\label{EQHypergeoInt2}
\int_0^{\hc} \omega_{rs}
=-\int_0^{\hc} d\tilde{c}\,\frac{\hb(\tilde{c})^{r-1} 
\tilde{c}^{s-1}}{\hb(\tilde{c})^4} 
= -\frac{1}{s} \hc^s\,
{}_2{\rm F}_1(\tfrac{s}{5},1-\tfrac{r}{5};1+\tfrac{s}{5}; \hc^5) \ .
\ee
In particular, the formula for the effective superpotential 
(\ref{superex}) follows directly from (\ref{EQHypergeoInt1}). Note
that the reference point $\hat{b}_0=0$ corresponds to $\hat{c}_0^5=1$,
and vice versa.

\subsubsection{Comparing different charts}

Since the differentials we have integrated are globally defined, the 
two expressions we obtain in different charts, namely
(\ref{EQHypergeoInt1}) and (\ref{EQHypergeoInt2}), must agree, once we
have taken into account that the lower bound of the integrals are
different. This can also be checked explicitly. In order to see this
we use the identity
\begin{align}
{}_2{\rm F}_1(\mfa,\mfb;\mfc;1-z) &=
\frac{\Gamma(\mfc)\Gamma(\mfa+\mfb-\mfc)}{\Gamma(\mfa)\Gamma(\mfb)}
{}_2{\rm   F}_1(\mfc-\mfa,\mfc-\mfb;
\mfc-\mfa-\mfb+1;z)\, z^{\mfc-\mfa-\mfb} \nonumber \\ 
&\quad + \frac{\Gamma(\mfc)\Gamma(\mfc-\mfa-\mfb)}
      {\Gamma(\mfc-\mfa)\Gamma(\mfc-\mfb)}
{}_2{\rm F}_1(\mfa,\mfb;\mfa+\mfb-\mfc+1;z) \ .
\end{align}
This allows us to rewrite the right hand side of
(\ref{EQHypergeoInt1}) as 
\begin{align} \label{zwischen}
&\frac{1}{r}\hb^r \hc^s\,
\frac{\Gamma(1+\tfrac{r}{5})\Gamma(-\tfrac{s}{5})}
{\Gamma(\tfrac{r}{5})\Gamma(1-\tfrac{s}{5})}\,
{}_2{\rm F}_1(1,\tfrac{r+s}{5};1+\tfrac{s}{5};\hc^5) 
+ \frac{1}{r} \hb^r\,
\frac{\Gamma(1+\tfrac{r}{5})\Gamma(\tfrac{s}{5})}
{\Gamma(\tfrac{r+s}{5})}\,
{}_2{\rm F}_1(\tfrac{r}{5};1-\tfrac{s}{5};1-\tfrac{s}{5};\hc^5) \ .
\end{align}
With the help of the identities
\begin{align}
{}_2{\rm F}_1(\mfa, \mfc;\mfc;z) &= (1-z)^{-\mfa}\\
{}_2{\rm F}_1(\mfa,\mfb;\mfc;z) &= 
(1-z)^{\mfc-\mfa-\mfb}\, {}_2{\rm F}_1(\mfc-\mfa,\mfc-\mfb;\mfc;z)
\end{align}
as well as properties of the $\Gamma$-function, (\ref{zwischen}) then 
becomes 
\be
-\frac{1}{s} \hc^s\,
{}_2{\rm F}_1(\tfrac{s}{5},1-\tfrac{r}{5};1+\tfrac{s}{5};\hc^5)
+ \frac{1}{r}\frac{\Gamma(1+\tfrac{r}{5})\Gamma(\tfrac{s}{5})}
{\Gamma(\tfrac{r+s}{5})} \ .
\ee
By the Gauss hypergeometric theorem the second term is precisely the 
value of the right hand side of (\ref{EQHypergeoInt1}) for
$\hat{b}=1$, while the first term agrees with
(\ref{EQHypergeoInt2}). Since $\hat{b}=1$ corresponds to $\hat{c}=0$,
the second term just accounts for the fact that the reference points
in the two line integrals (\ref{EQHypergeoInt1}) and
(\ref{EQHypergeoInt2}) are different, and we have therefore proven our
claim. In particular, this then implies that the function ${\cal W}$ 
defined by (\ref{superex}) solves both (\ref{Wb}) and (\ref{Wc}).

\bigskip

\centerline{\large \bf Acknowledgements}
\vskip .2cm

This research has been  partially supported by a TH-grant 
from ETH Z\"urich, the Swiss National Science Foundation and the Marie
Curie network `Constituents, Fundamental Forces and Symmetries of the
Universe' (MRTN-CT-2004-005104). The work of I.B. is supported by
an EURYI award.
We thank Carlo Angelantonj, Adel Bilal, Manfred Herbst, Hans Jockers, 
Bernard Julia, Elias Kiritsis, Wolfgang Lerche, Andreas Recknagel and
Stefan Theisen for useful discussions. We furthermore thank Christopher Beem for discussions and pointing out a typo in (3.14) that we corrected in version 2.

\end{document}